\documentclass[twocolumn,
secnumarabic, amssymb, amsmath, nofootinbib,tightenlines,showpacs,
nobibnotes, aps, prl]{revtex4}
\usepackage{amsfonts}
\usepackage{docs}%
\usepackage{bm}%
\usepackage{subfigure}
\usepackage[colorlinks=true,linkcolor=blue]{hyperref}%
\usepackage{graphicx}
\usepackage{dcolumn}

\begin{document}

\title{Topological Superfluid in one-dimensional Ultracold Atomic System\\ with Spin-Orbit Coupling}

\author{Zhongbo Yan}
\author{Xiaosen Yang}
\author{Shaolong Wan}
\email[]{slwan@ustc.edu.cn} \affiliation{Institute for Theoretical
Physics and Department of Modern Physics University of Science and
Technology of China, Hefei, 230026, \textbf{P. R. China}}
\date{\today}

\begin{abstract}
We propose a one-dimensional Hamiltonian $H_{1D}$ which supports
Majorana fermions when $d_{x^{2}-y^{2}}$-wave superfluid appears
in the ultracold atomic system and obtain the phase-separation
diagrams both for the time-reversal-invariant case and
time-reversal-symmetry-breaking case. From the phase-separation
diagrams, we find that the single Majorana fermions exist in the
topological superfluid region, and we can reach this region by
tuning the chemical potential $\mu$ and spin-orbit coupling
$\alpha_{R}$. Importantly, the spin-orbit coupling has realized in
ultracold atoms by the recent experimental achievement of
synthetic gauge field, therefore, our one-dimensional ultra-cold
atomic system described by $H_{1D}$ is a promising platform to
find the mysterious Majorana fermions.

\end{abstract}


\maketitle

\section{Introduction}

Since the discovery of the fractional quantum hall state \cite{D.
C. Tsui}, the concept of the topological order, which was first
proposed explicitly by Wen \cite{X. G. Wen}, has been developed
very fast and used in many condensed matter systems. A gapped
system, such as the Pfaffian state proposed by Moore and Read
\cite{G. Moore}, which possesses topological order, may have
practical use in the topological quantum computation (TQC) on
account of its quasi-particles'  non-trivial properties, such as
non-Abelian statistics, and the tolerant ability to the
decoherence from the environment \cite{A. Y. Kitaev1}.

Read and Green \cite{N. Read} pointed out that the zero energy
Majorana fermion modes exsiting at the cores of a 2D spinless
p-wave superconductor in the weak-phase \cite{S. Tewari, V.
Gurarie} are the same as the nonabelions in the Pfaffian state
\cite{G. Moore}, and they are non-Abelian quasi-particles \cite{D.
A. Ivanov}. Kitaev constructed a toy model and showed that
Majorana fermions exist as end states of a spin-polarized 1-D
superconductor \cite{A. Y. Kitaev2}. This model supplies an
insightful way to find the interesting single Majorana fermion.
Recently, many groups have proposed different systems to engineer
topological superconductivity (TSC) with Majorana fermions as a
bonus \cite{Liang Fu, Masatoshi Sato1, Xiao-Liang Qi1, Jay D. Sau,
Roman M. Lutchyn1, Yuval Oreg, Masatoshi Sato2, Shusa Deng}. Among
them, papers \cite{Roman M. Lutchyn1,Yuval Oreg} recognize that
the topological superconductivity can be perhaps most easily
engineered in one-dimensional semiconducting wires deposited on an
s-wave superconductor, and provide the first realistic
experimental setting for Kitaev's model and a platform to find and
manipulate single Majorana fermion by braiding \cite{J. Alicea}.
The authors of \cite{L. M. Wong} propose that Au wires in
proximity to doped LSCO($L_{2-x}Sr_{x}CuO_{4}$), a
($d_{x^{2}-y^{2}}$) wave superconductor can be a more promising
candidates for realizing single Majorana fermion.

In addition to fractional quantum hall systems and topological
superconductors, topological non-trivial superfluid, which can be
deduced from an underlying normal superfluid, e.g., s-wave
superfluid, also supports non-Abelian Majorana fermions. With the
rapidly developing technology available for the quantum control of
ultra-cold atomic systems, clean environment and highly tunable
parameters, ultracold atomic systems may serve as an idea platform
for the observation of topological superfluidity and topological
phase transition. Importantly, the realization of spin-orbit
coupling in ultracold atoms by the recent experimental achievement
of synthetic gauge field has made a firm step to engineer
topological superfluidity and non-Abelian quasi-particles therein
\cite{Y. -J. Lin, Y. -J. Lin1}.

With the introduction of the spin-orbit coupling, the energy gap
of superconductors or superfluids with asymmetry pairing wave
functions will close at some points in the first Brillouin zone.
If the system is under certain symmetries, e.g. time reversal
symmetry, particle-hole symmetry etc., and the manifold is not
closed, the system will possess robust gapless edge excitations
protected by these symmetries. However, once the symmetry
protecting the gapless excitations is broken, e.g. DIII class to D
in 3D case \cite{A. P. Schnyder}, the gapless excitations will no
longer be protected and will be absorbed by some random
impurities. Luckily, there are some cases \cite{A. P. Schnyder}
even the symmetry is broken, there will be gapless excitations
that can still exist robustly.

In this article, we study the one-dimensional ultracold atomic
system with spin-orbit coupling both for the time reversal
symmetry and the time-reversal-symmetry-breaking case. We find
that there is only one topologically nontrivial superfluid phase,
which is always protected by an energy gap away from the normal
superfluid, which agrees with the result in Ref.\cite{L. M. Wong}.
What most interests us is that we can reach the topological
superfluid (TSF) region explicitly by tuning the parameters
according to the phase-separation diagrams and directly probe the
single Majorana fermions.

\section{Model Study}

We consider a one-dimensional ultracold atomic system with
spin-orbit coupling and the Hamiltonian is
\begin{eqnarray}
H_{1D}=H_{t}+H_{SO}+H_{I}+H_{Z}, \label{1}
\end{eqnarray}
where
\begin{eqnarray}
&&H_{t}= -\frac{1}{2} t \sum_{j, \alpha} \left(
\psi_{j+1,\alpha}^{\dag}\psi_{j, \alpha} + h.c. \right) -
\sum_{j, \alpha} \mu \psi_{j, \alpha}^{\dag} \psi_{j, \alpha}, \nonumber\\
&&H_{SO}= -\frac{1}{2} \sum_{j, \alpha, \beta} \left( i \alpha_{R}
\psi_{j+1, \alpha}^{\dag}(\sigma_{y})_{\alpha\beta}
 \psi_{j, \beta} + h.c. \right), \nonumber\\
&&H_{I}= \frac{1}{2}\sum_{\alpha\beta}\sum_{ij}g_{ij}\psi_{i,\alpha}^{\dag}\psi_{j,\beta}^{\dag}\psi_{j,\beta}
 \psi_{i,\alpha} ,  \nonumber\\
&&H_{Z}= \sum_{j} V_{Z}
\left(\psi_{j,\uparrow}^{\dag}\psi_{j,\uparrow}-
\psi_{j,\downarrow}^{\dag}\psi_{j,\downarrow} \right), \label{2}
\end{eqnarray}
where $\psi_{j}$ is a fermion operator at site $j$, $\alpha$ and
$\beta$ are the spin indices, $t$ is the hopping amplitude, $\mu$
is the chemical potential, $\alpha_{R}$ is the spin-orbit coupling
strength, $H_{I}$ is the interaction, $g_{ij}$ is the interaction
strength between site $i$ and site $j$, and is negative in this
model, $\sigma_{y}$ is a pauli matrix, $H_{Z}$ is the Zeeman term
which breaks time reversal symmetry, and $V_{Z}$ denotes the
strength.

By Fourier transformation and mean field approach,
the above Hamiltonian will take the form
\begin{eqnarray}
H_{1D}=H_{t}+H_{SO}+H_{SF}+H_{Z}, \label{3}
\end{eqnarray}
where
\begin{eqnarray}
&&H_{t} = - \sum_{k,\alpha}\left[ t\cos(k) +\mu \right] \psi_{k,\alpha}^{\dag}\psi_{k,\alpha}, \nonumber\\
&&H_{SO} = -\sum_{k} i\alpha_{R}\sin(k) \left(
\psi_{k,\uparrow}^{\dag}\psi_{k,\downarrow}-
\psi_{k,\downarrow}^{\dag}\psi_{ k,\uparrow} \right), \nonumber\\
&&H_{SF} = \sum_{k} \left[ \Delta(k)
\psi_{k,\uparrow}^{\dag}\psi_{-k,\downarrow}^{\dag} \right. \nonumber\\
&& +\left.\Delta^{*}(k)\psi_{-k,\downarrow}
\psi_{k,\uparrow}\right]-|\Delta_{0}|^{2}/J , \nonumber\\
&&H_{Z} = \sum_{k}V_{Z} \left( \psi_{k,\uparrow}^{\dag}
\psi_{k,\uparrow} - \psi_{k,\downarrow}^{\dag} \psi_{k,\downarrow}
\right), \label{4}
\end{eqnarray}
where $\Delta(k) =\frac{1}{N}\sum_{k^{'}} g(k-k^{'})\langle
\psi_{-k^{'}, \downarrow} \psi_{k^{'}, \uparrow}
\rangle=\Delta_{0}\cos(k)$ is the pairing amplitude,
$g(k-k^{'})=g\cos(k-k^{'})$ is the Fourier form of $g_{ij}$, $g$
is a real constant. $\Delta_{0}$ is a complex constant, here we
make it real for convenience. Here the pairing we are interested
in is the $d_{x^{2}-y^{2}}$ type. $N$ is the number of sites and
the lattice constant $a$ is set as unit, $J=\frac{g}{N}$. Strictly
speaking, making mean field approximation is not proper for a
one-dimensional system, as fluctuations are strong and there is no
true long range order in a homogeneous system with non-zero
temperature in the thermodynamic limit, according to the
well-known Hohenberg-Mermin-Wagner theorem. However, as shown in
Ref.\cite{Xia-Ji Liu}, by confining the system in a  box with
finite length $L$ or in a harmonic trap to avoid this technology
difficulty, the authors found the mean-field methods provide a
useful description in the weakly or moderately interaction regimes
by comparing the mean-filed result with the exact of
asymptotically exact Bethe Ansatz solutions. In the following, we
will set the length of the system to be $L=Na=N$ and $T=0$

In the momentum space, under the Nambu spinor representation
$\Psi(k)^{\dag}=\{\psi_{k,\uparrow}^{\dag}, \psi_{ -k,
\downarrow}^{\dag}, \psi_{-k,\downarrow}, \psi_{k,\uparrow}\}$,
the Hamiltonian can be rewritten as
\begin{eqnarray}
H_{1D}=\frac{1}{2}\sum_{k} \Psi(k)^{\dag}H(k)\Psi(k)-
\Delta_{0}^{2}/J, \label{5}
\end{eqnarray}
where
\begin{eqnarray}
H(k) &=& \left[
\begin{array}{cc}
         h(k) & \Lambda(k) \\
         \Lambda(k)^{\dag} & -h^{T}(-k)
       \end{array}
       \right], \nonumber\\
h(k)&=& \varepsilon_{k} \sigma_{0} + \alpha_{R}\sin(k)
\sigma_{y}+V_{Z} \sigma_{z},
\nonumber\\
\Lambda(k) &=& i \Delta(k) \sigma_{y}. \label{6}
\end{eqnarray}
where $\varepsilon_{k}=-t\cos(k) -\mu$. After
diagonalizing, the Hamiltonian is of the form
\begin{widetext}
\begin{eqnarray}
H_{1D} = \frac{1}{2} \sum_{k} \left\{ \left[(E_{1}(k)-E_{3}(k)
\right] \alpha_{k,\uparrow}^{\dag} \alpha_{k,\uparrow} +
\left[E_{2}(k)-E_{4}(k)\right] \beta_{k,\downarrow}^{\dag}
\beta_{k,\downarrow}+ E_{3}(k)+
E_{4}(k) \right\}- \frac{\Delta_{0}^{2}}{J}+..., \label{7}
\end{eqnarray}
where $E_{1}(k), E_{2}(k), E_{3}(k), E_{4}(k)$ are in the order
$\{++, +-, -+, --\}$ of
\begin{eqnarray}
E(k)=\pm\sqrt{\varepsilon_{k}^{2}+\alpha_{R}^{2}\sin^{2}(k) +
\Delta_{0}^{2}\cos^{2}(k) +V_{Z}^{2}
\pm2\sqrt{\varepsilon_{k}^{2}\alpha_{R}^{2}\sin^{2}(k)
+\varepsilon_{k}^{2}V_{Z}^{2}+V_{Z}^{2}\Delta_{0}^{2}\cos^{2}(k)}},
\label{8}
\end{eqnarray}
\end{widetext} and ellipsis stands for the terms which are constant
and independent of $\Delta_{0}$, $\alpha_{k,\uparrow}^{\dag}
(\beta_{k,\downarrow}^{\dag})$ is the creation operators of the
excitation. The ground state is $\mid0\rangle$, which satisfies
$\alpha_{k,\uparrow}(\beta_{k,\downarrow})\mid0\rangle=0$, and
the energy of the ground state is
\begin{eqnarray}
E_{0} = \frac{1}{2} \sum_{k}\left[E_{3}(k)+E_{4}(k) \right]-\frac{\Delta_{0}^{2}}{J}, \label{9}
\end{eqnarray}
which needs to satisfy the condition of mean-field-approximation assumption
\begin{eqnarray}
\frac{2}{J} = \frac{1}{2}\sum_{k} \left[ \frac{A + V_{Z}^{2}}{E_{3}(k)A} +
\frac{A - V_{Z}^{2}}{E_{4}(k)A} \right] \cos^{2}(k), \label{10}
\end{eqnarray}
where $A=\sqrt{\varepsilon_{k}^{2}\alpha_{R}^{2}\sin^{2}(k) +
\varepsilon_{k}^{2}V_{Z}^{2} + V_{Z}^{2}
\Delta_{0}^{2}\cos^{2}(k)}$.

\section{Results and Discussions}

First, for the time-reversal invariant case $(V_{Z}=0)$, we have
\begin{eqnarray}
E_{3}(k), E_{4}(k)=-\sqrt{(\varepsilon_{k} \pm
\alpha_{R}\sin(k))^{2} + \Delta_{0}^{2}\cos^{2}(k)}. \label{11}
\end{eqnarray}
When $V_{Z}=0$, the Hamiltonian $H(k)$ possesses a chiral
symmetry. Therefore, the Hamiltonian can be brought into an
off-diagonal form by a unitary transformation \cite{Xiao-Liang
Qi2}
\begin{eqnarray}
\tilde{H}&=& VH(k)V^{\dag}=\left(
\begin{array}{cc}
          & h(k)+i\mathcal{T}\Lambda(k)^{\dag} \\
         h(k)-i\mathcal{T}\Lambda(k)^{\dag} &
       \end{array}
       \right)  \nonumber \\
       &\simeq&\left(
\begin{array}{cc}
          & Q_{k} \\
          Q_{k}^{\dag}&
       \end{array}
       \right) , \nonumber\\
\label{12}
\end{eqnarray}
with
\begin{eqnarray}
V=\frac{1}{\sqrt{2}}\left(
\begin{array}{cc}
          1& -1 \\
         1 & 1
       \end{array}
       \right)\left(
\begin{array}{cc}
          1&  \\
          &-i\mathcal{T}
       \end{array}
       \right), \nonumber\\
\label{13}
\end{eqnarray}
and $\mathcal{T}=i\sigma_{y}$,
$Q_{k}=\frac{1}{2}[e^{i\theta_{-}(k)}
(\sigma_{0}-\sigma_{y})+e^{i\theta_{+}(k)}(\sigma_{0}+\sigma_{y})]$,
where $e^{i\theta_{\pm}(k)}=\frac{-t\cos(k)-\mu \pm
\alpha_{R}\sin(k)+ i\Delta_{0}\cos(k)}{\sqrt{[-t\cos(k)-\mu \pm
\alpha_{R}\sin(k)]^{2}+ [\Delta_{0}\cos(k)]^{2}}}$. In
Eq.(\ref{12}), the meaning of $\simeq$ is the magnitude of the
eigenvalues of $\tilde{H}$ has normalized.

In one dimension, the Fermi-surface topological invariant (FSTI)
of a time-reversal-invariant (TRI) superconductor or superfluid is
\cite{Xiao-Liang Qi2}
\begin{eqnarray}
N_{1D}=\prod_{s}[sgn(\delta_{s})],\label{14}
\end{eqnarray}
where $s$ is summed over all the Fermi points between $0$ and $\pi$.
If a system has odd number of Fermi points between $0$ and $\pi$ with
negative pairing, in other words, $N_{1D}=-1$, the system is non-trivial,
otherwise $N_{1D}=1$, and the system is trivial.

In our system, there is only one Fermi point ($k=\frac{\pi}{2}$
with $|\mu|=\alpha_{R}$) between $0$ and $\pi$, so the sign of
$\delta_{s}$ directly determines the Hamiltonian is trivial or
non-trivial. The sigh of $\delta_{s}$ is positive or negative
corresponding to the change in $\theta_{\pm}(k)$ across
$k=\frac{\pi}{2}$ is $-\pi$ or $\pi$ \cite{Xiao-Liang Qi2}. Here,
in the weak pairing limit, by increasing $k$ from
$\frac{\pi}{2}-\epsilon$ to $\frac{\pi}{2}+\epsilon$ with
$\alpha_{R}\in(|\mu|-\delta,|\mu|+\delta)$ ($\epsilon$, $\delta$
are small positive constants), we find, for $\alpha_{R}>\mu$ (we
only consider $\mu>0$ on account of symmetry), the real and
imaginary part of $-t\cos(k)-\mu
+\alpha_{R}\sin{k}+i\Delta_{0}\cos(k)$ change as
\begin{eqnarray}
-\tilde{\delta}+i\tilde{\epsilon}\longrightarrow
\tilde{\delta}+i\tilde{\epsilon}\longrightarrow
\tilde{\delta}-i\tilde{\epsilon}
\label{15}
\end{eqnarray}
with
\begin{eqnarray}
\theta_{+}(k)\longrightarrow&\theta_{+}(k)-\pi&\longrightarrow\theta_{+}(k)+2\pi, \nonumber \\
&\Delta\theta_{+}(k)=+\pi&, \label{16}
\end{eqnarray}
for $\alpha_{R}<\mu$,
\begin{eqnarray}
-\tilde{\delta}+i\tilde{\epsilon}\longrightarrow
-\tilde{\delta}-i\tilde{\epsilon}\longrightarrow
\tilde{\delta}-i\tilde{\epsilon}
\label{17}
\end{eqnarray}
with
\begin{eqnarray}
\theta_{+}(k)\longrightarrow&\theta_{+}(k)-2\pi&\longrightarrow\theta_{+}(k)+\pi, \nonumber \\
&\Delta\theta_{+}(k)=-\pi&, \label{18}
\end{eqnarray}
($\tilde{\delta},\tilde{\epsilon}$ are small positive real
numbers) therefore, when $\mu<\alpha_{R}$,
$N_{1D}=sgn(\delta_{s})=-1$, the system is non-trivial; when
$\mu>\alpha_{R}$, $N_{1D}=sgn(\delta_{s})=1$, the system is
trivial.

\begin{figure}
\subfigure{\includegraphics[width=4cm,height=4cm]{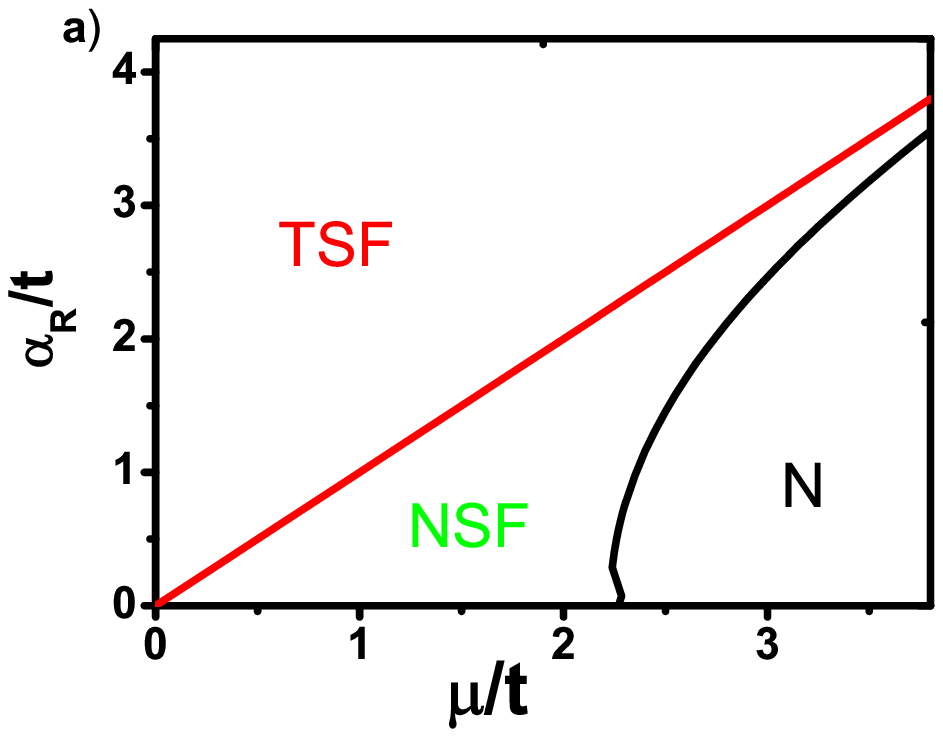}}\label{Fig.1a}
\subfigure{\includegraphics[width=4cm,height=4cm]{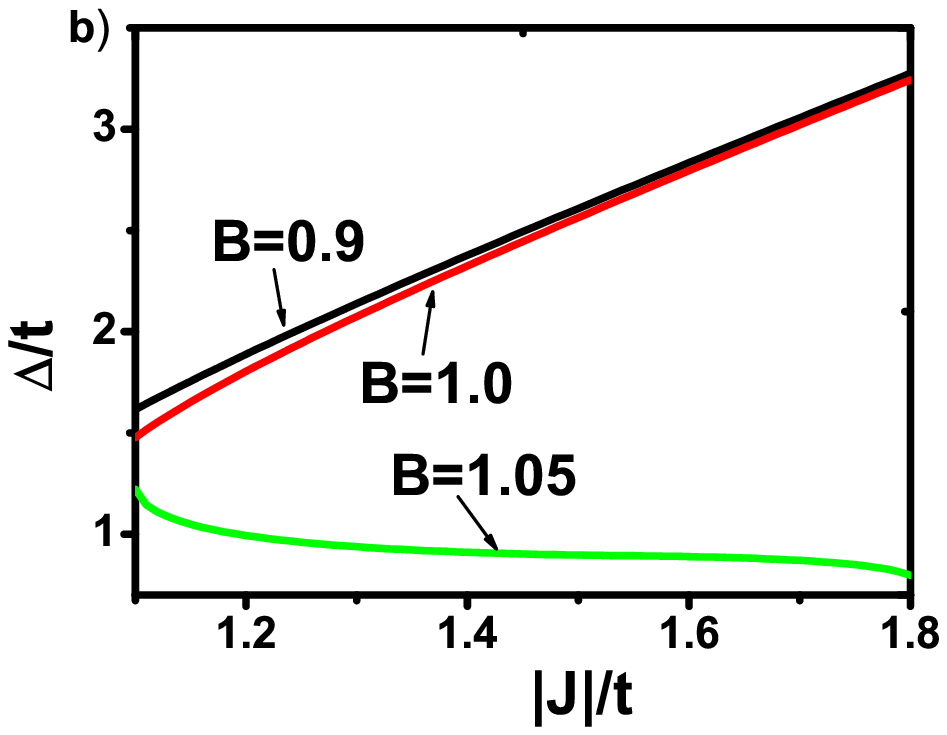}}\label{Fig.1b}
\subfigure{\includegraphics[width=4cm,height=4cm]{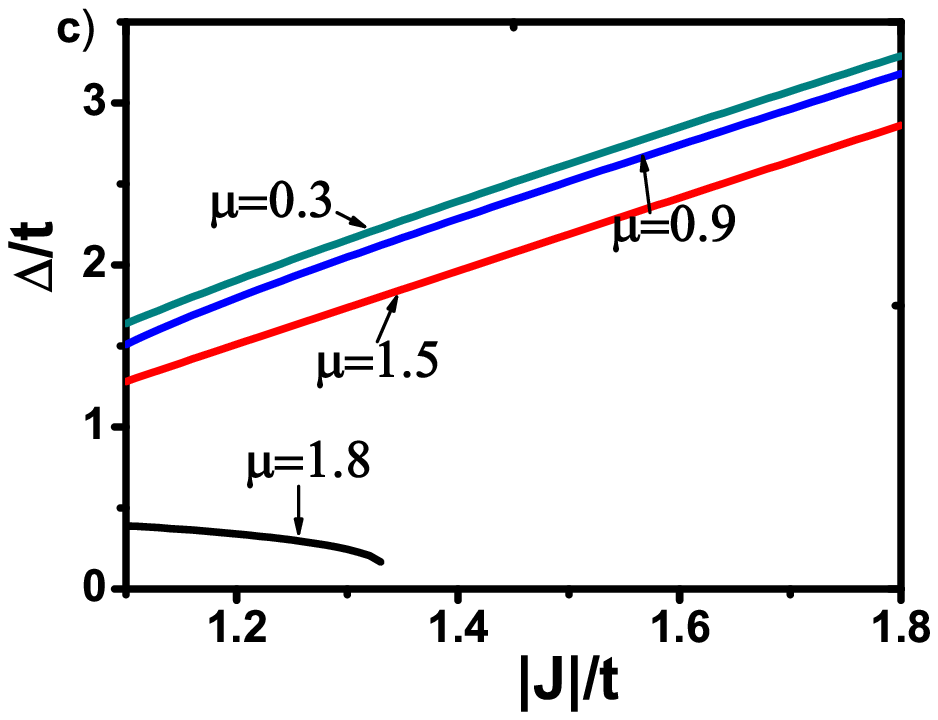}}\label{Fig.1c}
\subfigure{\includegraphics[width=4cm,height=4cm]{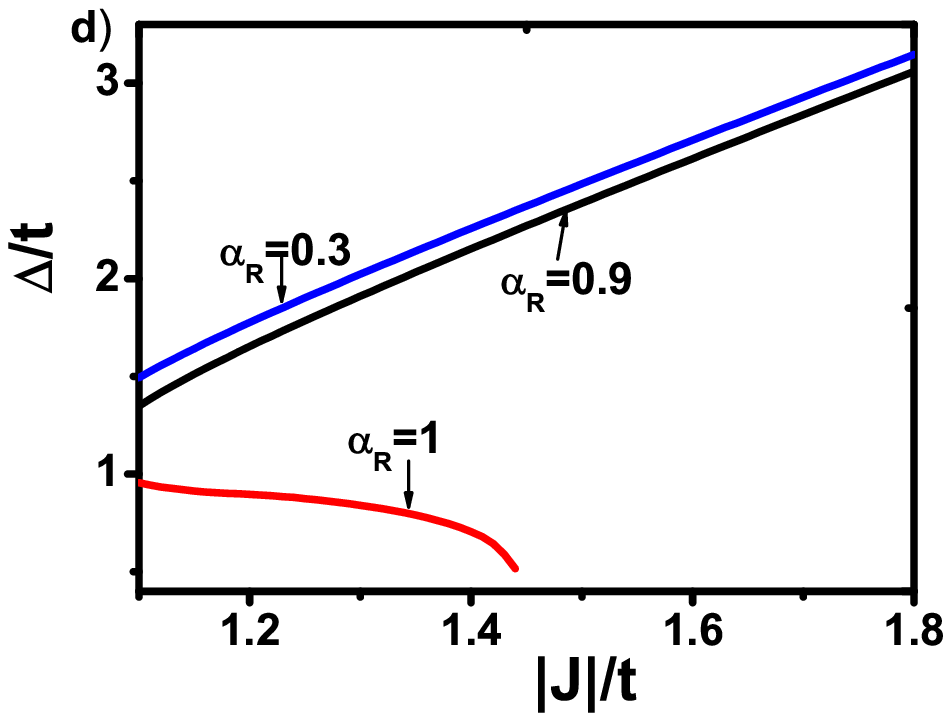}}\label{Fig.1d}
\caption{(Color online) (a) The parameters of $H_{1D}$ are:
$J=-1.2, V_{Z}=0, t=1$, there are two critical lines, which
separate the normal superfluid (NSF) from the topological
superfluid (TSF) and normal state (N) respectively. (b) $t=1,
\mu=0.45, \alpha_{R}=0.3$, changing the Zeeman field $V_{Z}$ from
0.9 to 1.05, we see $\Delta_{0}$ monotonically decreases with the
increasing $V_{Z}$, and when the Zeeman field is strong enough,
$\Delta_{0}$ will be dramatically suppressed and no longer
monotonically increase with $J$, the attractive interaction
strength. (c) $t=1, V_{Z}=0.9, \alpha_{R}=0.3$; (d) $t=1,
V_{Z}=0.9, \mu=0.45$, (c), (d) are quite like (b), $\Delta_{0}$
also monotonically decreases with the increasing $\mu$ and
$\alpha_{R}$, and there are also critical value $\mu_{c}$ and
$\alpha_{R_{c}}$, which dramatically suppress $\Delta_{0}$.}
\label{Fig.1}
\end{figure}

Base on Eq.(\ref{10}), Eq.(\ref{11}) and Eq.(\ref{14}) and the
above analysis, if we fix the strength of spin-orbit coupling, we
find that there exists a critical chemical potential $\mu_{c}$
where a superfluid-normal state transition takes place, as shown
in Fig.\ref{Fig.1}(a). Above the critical value, it's the normal
state where the mean-field-approximation assumption is not valid.
Below it, the superfluid phase appears and exists only when the
chemical potential is not too large. In this region, when
$\mu>\alpha_{R}$, the $Z_2$ invariant $N_{1D}=1$ where is the
normal superfluid phase region, and when $\mu<\alpha_{R}$,
$N_{1D}=-1$ where is the topological superfluid phase region, as
shown in Fig.\ref{Fig.1}(a). The line $\mu=\alpha_{R}$ separating
NSF from TSF is a critical line, crossing this line, the
topological number changes and the topological phase transition
takes place. Because there is no symmetry breaking while crossing
the critical line, the topological order $N_{1D}$ is the only
parameter to distinguish these two phases.

Second, for the time-reversal-symmetry-breaking case $(V_{Z} \neq
0)$, the $Z_{2}$ Majorana number $\mathcal{M}$ of $H(k)$ is the
new topological invariant to determine the system is topologically trivial
or topologically non-trivial. Following the Refs.\cite{A. Y. Kitaev2, L. M. Wong, Roman M. Lutchyn2}
the Majorana number can be obtained by the formula
\begin{eqnarray}
\mathcal{M}=sgn[PfB(0)]sgn[PfB(\pi)]=\pm1, \label{19}
\end{eqnarray}
where $\pm1$ corresponds to topologically trivial and non-trivial
states and the antisymmetric matrix $B(k)$ is defined as
$B(k)=H_{1D}(\sigma_{x}\otimes\sigma_{0})$. In terms of the
parameters of the Hamiltonian, the Majorana number can be written
as
\begin{eqnarray}
sgn{[(t+\mu)^{2}-(V_{Z}^{2}-\Delta_{0}^{2})]
[(-t+\mu)^{2}-(V_{Z}^{2}-\Delta_{0}^{2})]}, \nonumber \\
\label{20}
\end{eqnarray}
so the Majorana number is $\mathcal{M}=-1$ when $|\sqrt{V_{Z}^{2}-\Delta_{0}^{2}}-t|
< |\mu|<|\sqrt{V_{Z}^{2}-\Delta_{0}^{2}}+t|$ and $\mathcal{M}=1$ otherwise.

In this case, we find that the phase diagrams will have much more
structures than the case discussed above. We find that the pairing
amplitude $\Delta_{0}$ monotonically decreases with the increasing
$V_{Z}$, and when the Zeeman field is strong enough, $\Delta_{0}$
will be dramatically suppressed and no longer monotonically
increase with $|J|$, the attractive interaction strength, as shown
in Fig.\ref{Fig.1}(b). This can be explained by the fact that with
the Zeeman field increasing, the polarization becomes stronger and
stronger, and finally, this will destroy the singlet
$d_{x^{2}-y^{2}}$ pairing. We also find that $\mu$ and
$\alpha_{R}$ have the same effects on the pairing amplitude as the
Zeeman field, as shown in Fig.\ref{Fig.1}(c)(d).

\begin{figure}
\subfigure{\includegraphics[width=4cm,height=4cm]{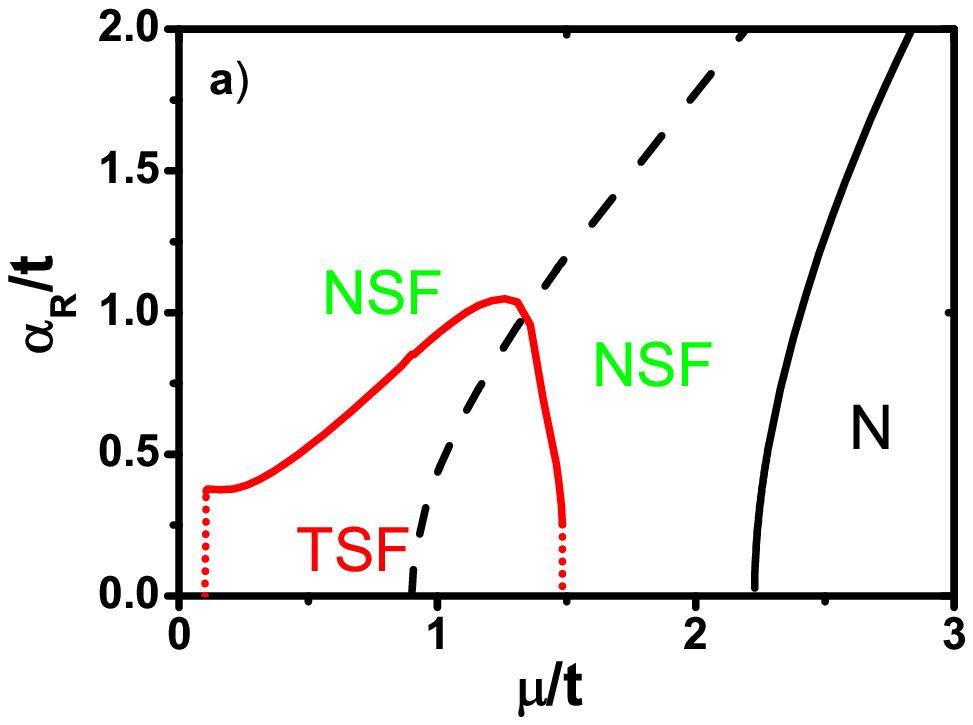}}\label{Fig.2a}
\subfigure{\includegraphics[width=4cm,height=4cm]{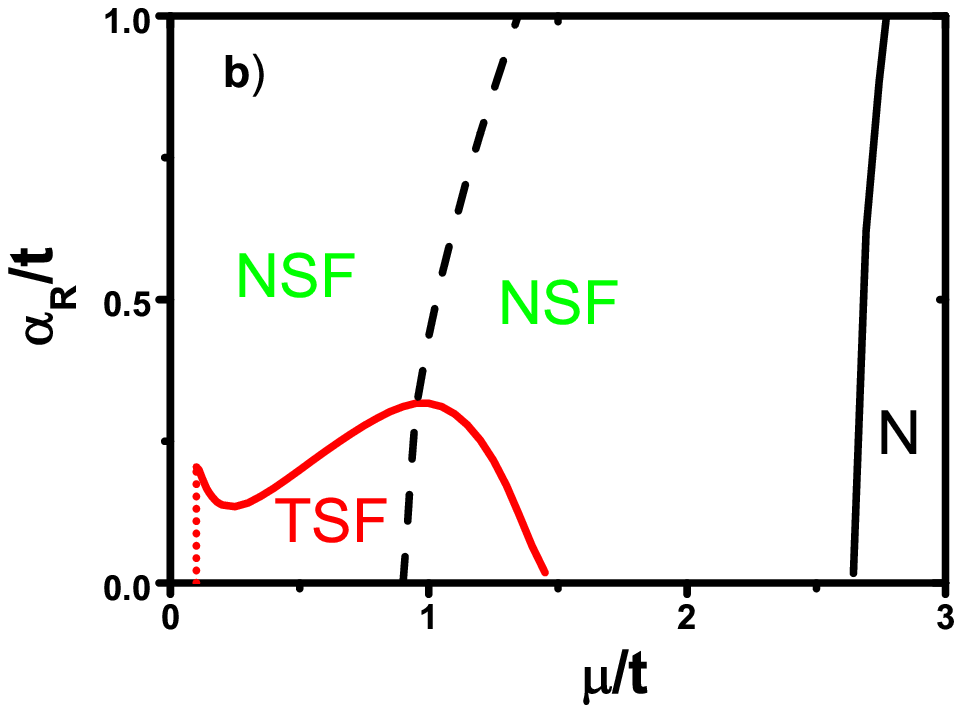}}\label{Fig.2b}
\subfigure{\includegraphics[width=4cm,height=4cm]{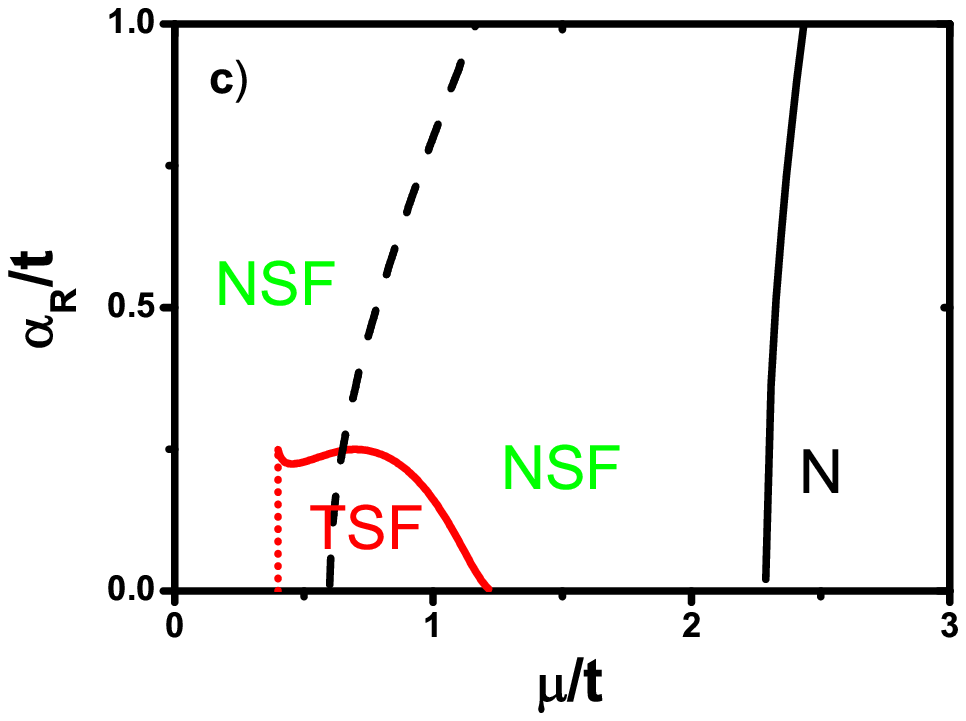}}\label{Fig.2c}
\subfigure{\includegraphics[width=4cm,height=4cm]{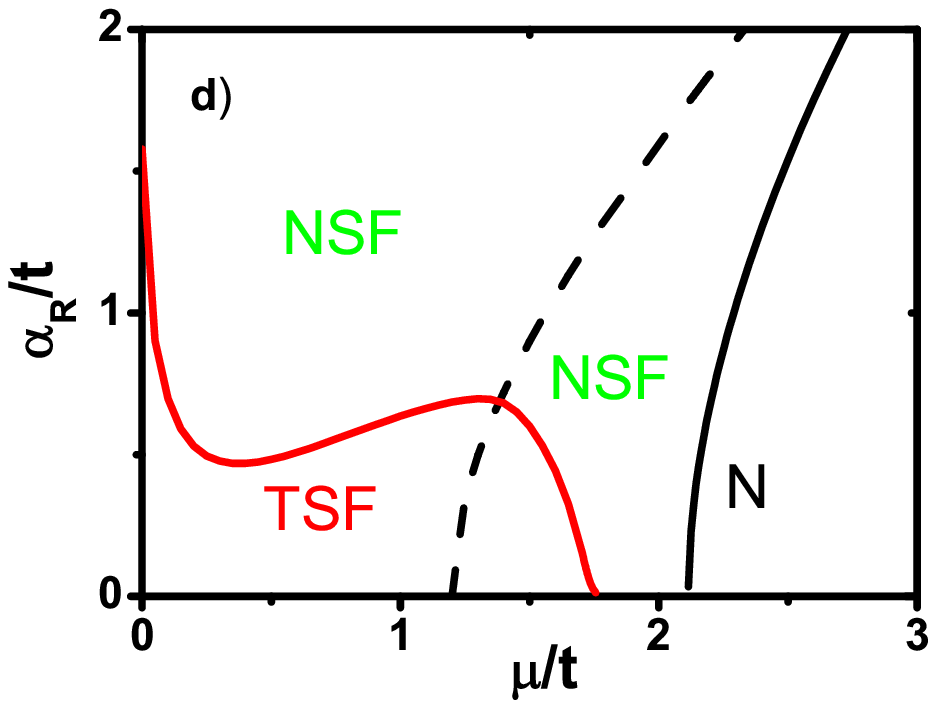}}\label{Fig.2d}
\caption{(Color online) The parameters of the Hamiltonian are: (a)
$t=1, V_{Z}=0.9, J=-1.2;$ (b) $t=1, V_{Z}=0.9, J=-1.5;$ (c) $t=1,
V_{Z}=0.6, J=-1.2;$ (d) $t=1, V_{Z}=1.2, J=-1.6.$ the red solid
line is a critical line, separating TSF from NSF, the dotted line
is a natural extension of the red solid line, it is also a phase
boundary, separating TSF from NSF; the black solid line is also a
critical line, separating NSF from N. The dashed line is not a
critical line, across it, no phase transition takes place.
Comparing (b)(c) with (a), we find that the TSF regions shrink
either with $V_{Z}$ decreasing or with $\mid J\mid$ increasing. In
(d), we fix the ration $V_{Z}/J$ to be the same as (a) and keep
$t=1$, then we find the TSF region is broadened.} \label{Fig.2}
\end{figure}

In the following, we will discuss the most interesting part of our
work. According to the phase diagram Fig.\ref{Fig.2}(a) and
Eq.(\ref{20}), there is a critical line (red solid line),
corresponding to $\mu= \sqrt{V_{Z}^{2}-\Delta_{0}^{2}}+t$ and
$\mu= t-\sqrt{V_{Z}^{2}-\Delta_{0}^{2}}$, that separates TSF from
NSF. Here TSF means that single Majorana fermion exists in this
region, while NSF means superfluid with Majorana doublets (as
$V_{Z}<\sqrt{t^{2}+\Delta_{0}^{2}}$), which are no longer
topologically protected due to time-reversal symmetry breaking.
There is also a critical line (black solid line), corresponding to
$\Delta_{0}=0$, that separates NSF from N. In Fig.\ref{Fig.2}(a),
there is a dashed line. This dashed line corresponds to
$\mu=\sqrt{V_{Z}^{2}+\alpha_{R}^{2}}$. On this line, the energy
gap closes, however, from Fig.\ref{Fig.2}(a), we see the Majorana
numbers on both sides of this line are equivalent, either both are
trivial or both are non-trivial. Therefore, there is no
topological phase transition across this line. According to
Fig.\ref{Fig.2}(a), we can tune $\mu/t$ to $(0.1,1.45)$,
$\alpha_{R}/t$ to $(0,1)$ and reach the TSF region, then we can
apply the spatially resolved radio-frequency spectroscopy, which
would show a well isolated signal at zero energy, to detect the
associated single Majorana fermions.

In order to obtain the information that how the TSF region changes
with other tunable parameters and make single Majorana fermion
measurements more achievable, Figs.\ref{Fig.2}(b)-(d) are given.
Among them, Fig.\ref{Fig.2}(b) and Fig.\ref{Fig.2}(c) show that by
increasing the attractive interaction or reducing the Zeeman
field, although the NSF region are broadened, the TSF region are
greatly reduced. However, by increasing both the attractive
interaction and the Zeeman field, as shown in Fig.\ref{Fig.2}(d),
the TSF region are greatly broadened. Such effect will be quite
useful in experiments, as the larger the TSF region is, the more
detectable the single Majorana fermions are. From
Figs.\ref{Fig.2}(a)(b)(d), we can find TSF tends to form in the
high polarization area where the pairing amplitude is small and
the imbalance of the chemical potential is large.

\section{Conclusions}

In this paper, we propose the one-dimensional Hamiltonian $H_{1D}$
and discuss it both for the time-reversal-invariant case and the
time-reversal-symmetry-breaking case.  By numerical solving the
self-consistent equation Eq.(\ref{10}), we obtain different
phase-separation diagrams under different conditions. From the
diagrams, we find, with the spin-orbit coupling and the Zeeman
field, TSF exists. By tuning the parameters, such as $\mu$ and
$\alpha_{R}$, we can reach the TSF region, where single Majorana
fermions exist. To detect single Majorana fermions, we can apply
the spatially resolved radio-frequency spectroscopy, which would
show a well isolated signal at zero energy.

With the rapidly developing technology available for the quantum
control and the introduction of spin-orbit coupling to ultra-cold
atomic systems, we believe that our one-dimensional ultra-cold
atomic system described by $H_{1D}$ is a promising platform to
find the mysterious Majorana fermions.

\begin{acknowledgments}
This work is supported by NSFC Grant No.10675108.
\end{acknowledgments}

\end{document}